\documentclass[letterpaper]{article} 
\usepackage{aaai2026}  
\usepackage{times}  
\usepackage{helvet}  
\usepackage{courier}  
\usepackage[hyphens]{url}  
\usepackage{graphicx} 
\usepackage{natbib}  
\usepackage{caption} 
\usepackage{algorithm}
\usepackage{algorithmic}
\usepackage{amssymb}
\usepackage{utfsym}
\usepackage{amsmath}
\usepackage{multirow}
\usepackage{booktabs}
\usepackage{subfig}
\newtheorem{theorem}{Theorem}

\usepackage{newfloat}
\usepackage{listings}
\DeclareCaptionStyle{ruled}{labelfont=normalfont,labelsep=colon,strut=off} 
\floatstyle{ruled}
\newfloat{listing}{tb}{lst}{}
\floatname{listing}{Listing}
\title{AdaField: Generalizable Surface Pressure Modeling with Physics-Informed Pre-training and Flow-Conditioned Adaptation}
\author{
	Junhong Zou\textsuperscript{\rm 1,2},
	Wei Qiu\textsuperscript{\rm 2,3}, 
	Zhenxu Sun\textsuperscript{\rm 3}, 
	Xiaomei Zhang\textsuperscript{\rm 1,2}, 
	Zhaoxiang Zhang\textsuperscript{\rm 1}, 
	Xiangyu Zhu\textsuperscript{\rm 1,2}\thanks{Corresponding authors}
}
\affiliations{
	\textsuperscript{\rm 1} MAIS, Institute of Automation, Chinese Academy of Sciences, Beijing, China,\\
	\textsuperscript{\rm 2} School of Artificial Intelligence, University of Chinese Academy of Sciences, Beijing, China,\\
	\textsuperscript{\rm 3} Key Laboratory for Mechanics in Fluid Solid Coupling Systems, Institute of Mechanics, Chinese Academy of Sciences, Beijing, China \\
	zoujunhong2022@ia.ac.cn, xiangyu.zhu@ia.ac.cn
}

\usepackage{bibentry}

\begin{document}

\maketitle

\begin{abstract}
The surface pressure field of transportation systems, including cars, trains, and aircraft, is critical for aerodynamic analysis and design. In recent years, deep neural networks have emerged as promising and efficient methods for modeling surface pressure field, being alternatives to computationally expensive CFD simulations. Currently, large-scale public datasets are available for domains such as automotive aerodynamics. However, in many specialized areas, such as high-speed trains, data scarcity remains a fundamental challenge in aerodynamic modeling, severely limiting the effectiveness of standard neural network approaches. To address this limitation, we propose the Adaptive Field Learning Framework (AdaField), which pre-trains the model on public large-scale datasets to improve generalization in sub-domains with limited data. AdaField comprises two key components. First, we design the Semantic Aggregation Point Transformer (SAPT) as a high-performance backbone that efficiently handles large-scale point clouds for surface pressure prediction. Second, regarding the substantial differences in flow conditions and geometric scales across different aerodynamic subdomains, we propose Flow-Conditioned Adapter (FCA) and Physics-Informed Data Augmentation (PIDA). FCA enables the model to flexibly adapt to different flow conditions with a small set of trainable parameters, while PIDA expands the training data distribution to better cover variations in object scale and velocity. Our experiments show that AdaField achieves SOTA performance on the DrivAerNet++ dataset and can be effectively transferred to train and aircraft scenarios with minimal fine-tuning. These results highlight AdaField’s potential as a generalizable and transferable solution for surface pressure field modeling, supporting efficient aerodynamic design across a wide range of transportation systems.
\end{abstract}

\begin{links}
     \link{Code}{https://github.com/zoujunhong/UniField}
\end{links}

\begin{figure}[t!]
	\centering
	\includegraphics[width=0.95 \columnwidth]{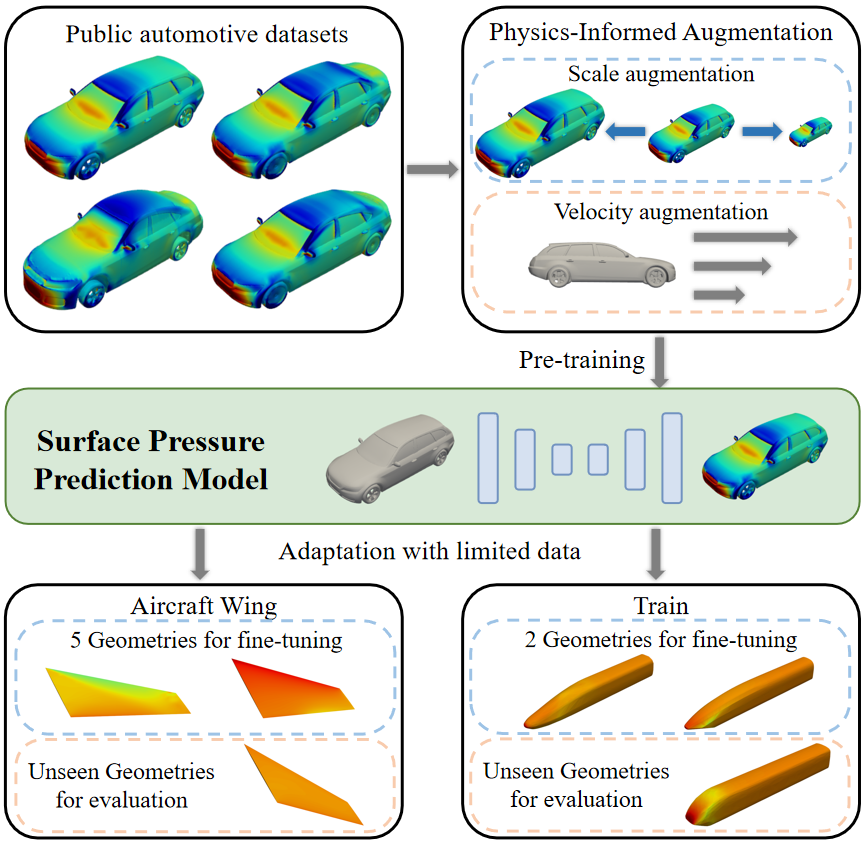}
	\caption{Overview. AdaField augments the public automotive dataset for model pre-training and then fine-tunes for adapting to other fields with limited data, such as aircraft wings and trains.}
	\label{fig: summary}
\end{figure}

\section{Introduction}

In the aerodynamic analysis of transportation systems, including cars, trains, and aircrafts, the surface pressure fields plays a fundamental role in determining key physical quantities such as lift, drag, lateral force, and overturning moment \cite{qiu2025compressed}. Accurately capturing high-resolution surface pressure fields is essential for optimizing aerodynamic performance, ensuring structural integrity, and improving fuel efficiency, among other practical concerns \cite{gao2025accurate}. Traditionally, surface pressure field simulations are conducted through computational fluid dynamics (CFD), which is expensive, time-consuming, and struggle to scale across the vast design space of shapes and flow conditions encountered in modern transportation systems.

Recently, neural networks have emerged as promising and efficient alternatives for surface pressure modeling, benefiting from their capacity to learn complex patterns from large datasets \cite{TripNet,FIGConvNet,Wang_Zhang_Liu_Zhao_Lin_Chen_2025,DrivAerNet++}. Compared with CFD, which usually requires hundreds of CPU hours to simulate the pressure field, neural networks can typically complete the prediction within one second. However, data scarcity remains a fundamental obstacle to deploying these models broadly in aerodynamic applications. Although extensive public datasets are available for some domains, most notably automotive aerodynamics \cite{DrivAerNet++, ashton2024drivaer, ashton2024ahmed, ashton2024windsor}, many specialized subfields, such as trains or aircrafts, lack sufficient data. This limitation substantially limits the effectiveness of conventional neural network approaches.
Moreover, substantial differences in geometric scale and flow conditions across transportation modalities further complicate model generalization and transfer. For instance, cars typically use driving speed and wind speed as inputs, while aircraft \cite{Wang_Zhang_Liu_Zhao_Lin_Chen_2025} often use Mach number (Ma) and angle of attack (AoA), and high-speed trains operate at distinct velocity and geometric scales against cars. As a result, models pre-trained in one domain often perform poorly when applied directly to others, highlighting the urgent need for approaches that enable efficient knowledge transfer and generalization under data scarcity.

To address these challenges, we propose the Adaptive Field Learning Framework (AdaField), which focuses on pre-training models on public automotive datasets and adapt them to other data-limited sub-domains, as shown in Figure \ref{fig: summary}. AdaField first introduces Semantic Aggregation Point Transformer (SAPT), which serves as a high-performance backbone, integrating local vector self-attention adapted from Point Transformer \cite{zhao2021pointtransformer} and a semantic aggregation module adapted from Slot Attention \cite{locatello2020object} to efficiently extract geometric features. Local vector self-attention restricts attention computation to the k-nearest neighbors, thus reducing memory requirement for processing point clouds with a large number of points. The semantic aggregation module learns to decompose point clouds into parts, capturing structural commonalities across complex surfaces. On the other hand, to improve cross-domain generalization and address the substantial differences in flow conditions and geometric scales among various transportation systems, AdaField introduces two complementary strategies. First, the Flow-Conditioned Adapter (FCA), designed based on adapter \cite{houlsby2019parameter}, injects flow-specific information into the model through a lightweight set of trainable parameters, enabling efficient fine-tuning for new aerodynamic regimes. Second, we employ a Physics-Informed Data Augmentation (PIDA) strategy grounded in the Navier–Stokes equations \cite{constantin1988navier}, expanding the training distribution across diverse object scales and velocities. Together, FCA and PIDA enable AdaField to achieve robust transferability and generalization under data-scarce conditions.

We validate the effectiveness of AdaField through experiments on DrivAerNet++ \cite{DrivAerNet++}, a publicly available large-scale CFD dataset for automotive aerodynamics. Our evaluation focuses on the prediction of surface pressure fields, where AdaField consistently achieves SOTA performance across multiple standard metrics.
To further demonstrate its cross-domain generalization capability, we evaluate AdaField on two additional datasets representing distinct transportation systems: a train dataset \cite{qiu2025compressed} and an aircraft wing dataset \cite{Wang_Zhang_Liu_Zhao_Lin_Chen_2025}. In these experiments, AdaField is first pre-trained on DrivAerNet++, and then fine-tuned with only a small number of samples from each target domain. The results show that AdaField exhibits strong few-shot transferability and generalization to novel aerodynamic scenarios, underscoring its practical potential for data-scarce subdomains beyond automotive applications.

To summarize, our main contributions are as follows:
\begin{itemize}
	\item We propose AdaField, a framework for predicting the pressure field on the surface of transportation systems. By integrating local vector self-attention with a semantic aggregation module, it can efficiently handle point clouds with a large number of points and maintain high-performance predictions.
	
	\item We develop FCA to facilitate domain adaptation under substantial differences in flow conditions across transportation systems, as well as PIDA, a data augmentation strategy which expands the diversity of training data in terms of object scale and velocity, enabling generalization to unseen aerodynamic conditions.
	
	\item We achieve SOTA performance on DrivAerNet++, and also conduct comprehensive cross-domain generalization studies, demonstrating that AdaField, when pre-trained on large-scale automotive data, can be effectively adapted to train and aircraft scenarios with minimal fine-tuning, verifying its practical potential as a solution for aerodynamic design tasks in data-scarce settings.
	
\end{itemize}

\section{Related Works}

\subsection{Surface Pressure Prediction}
Surface pressure prediction aims to infer the pressure distribution across a vehicle’s surface given its geometry and flow conditions. Traditional CFD software simulates surface fields by numerically solving the governing equations of fluid dynamics, such as the Navier–Stokes equations, over a discretized computational domain, but is computationally heavy, motivating neural networks as alternatives. DrivAerNet++ \cite{DrivAerNet++} is a large-scale aerodynamic dataset. It provides thousands of car geometries, CFD flow and pressure fields, parametric models, and aerodynamic coefficients, fostering robust training and evaluation for automotive aerodynamics. It also introduce RegDCGNN \cite{Elrefaie_2025}, a dynamic graph convolutional neural network to regress aerodynamic parameters, while avoiding the overhead of rendering or SDF preprocessing.
In the subsequent work, Transolver \cite{Transolver,luo2025transolveraccurateneuralsolver} leverages a Transformer-based PDE solver with a physics-inspired slice attention that groups mesh points into learnable physical-state slices, enabling scalable generalization across complex geometries. 
Factorized Implicit Global Convolution (FIGConvNet) \cite{FIGConvNet} efficiently learns global interactions across 3D meshes via implicit factorization, reducing complexity while preserving accuracy.
TripNet \cite{TripNet} encodes 3D car geometry into compact triplane representations, enabling point-wise predictions of pressure and full flow fields.

\begin{figure*}[t]
	\centering
	\includegraphics[width=2.1\columnwidth]{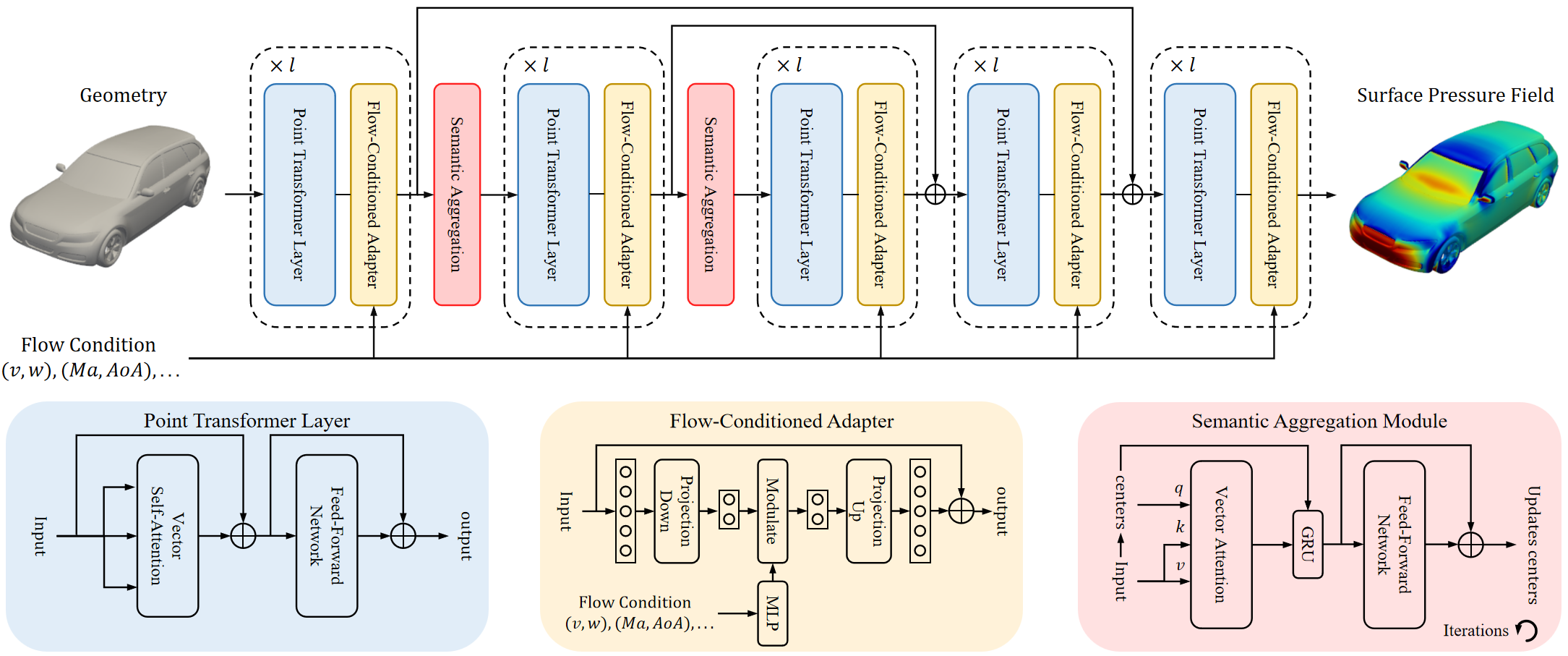}
	\caption{Architecture. The geometry represented by point clouds, along with flow conditions such as velocity, Mach number, and angle of attack, serve as the input. The model performs feature extraction through the Point Transformer layer and the Flow-Conditioned Adapter. The overall network structure adopts the U-Net style, with the downsampling process using a semantic aggregation module to aggregate point clouds through feature guidance. The upsampling process employs KNN interpolation.}
	\label{fig: architecture}
\end{figure*}

\subsection{Physics‑Informed Model Optimization}

Incorporating physical laws into neural network loss functions through physics-informed constraints or regularization terms enhances model generalizability, improves interpretability, and ensures physical consistency in predictions, particularly valuable for aerodynamic applications where traditional data-driven models might violate fundamental aerodynamic principles.
Physics‑Informed Neural Networks (PINNs) \cite{RAISSI2019686} are typical scheme to enforce governing PDEs or conservation laws via composite loss functions, which is a common practice in field modeling \cite{Yan_2024,MAO2020112789,10.1115/1.4050542}. 
Transolver \cite{Transolver} propose physics-attention to embed physics-aware slicing inside attention modules, improving performance on complex geometries by aligning latent tokens with physical states across mesh domains.
PEINR \cite{shen2025peinr} decouples temporal and spatial components through Gaussian temporal encoding, enhancing high-dimensional features and nonlinear characteristics in temporal information, and localized spatial encoding.
Recent PDE-Solvers \cite{zhou2024unisolverpdeconditionaltransformersuniversal} propose to leverage LLM for representing PDE symbols, such as coefficients, boundary conditions, e.t.c, as conditions in Transformer forward process, enabling strong generalization across diverse physics, advancing physics-aware modeling capabilities.
Wang et al. \shortcite{Wang_Zhang_Liu_Zhao_Lin_Chen_2025} propose to incorporate flow condition via Cross-Attention Fusion and achieve Physical‑Informed training by introducing Physical Equations derived from flow conditions (e.g. Mach number, angle of attack) to construct losses as physics‑informed regularization.

\section{Method}

Our network framework, AdaField, is illustrated in Figure \ref{fig: architecture}. The inputs to the network consist of geometric shape and flow conditions. The geometry is represented as a point cloud $G \in \mathbb{R}^{N \times 3}$, where $N$ denotes the number of surface nodes. The flow conditions are encoded as a vector $C \in \mathbb{R}^{D_f}$, where $D_f$ is the number of considered flow parameters. The network predicts a scalar pressure coefficient for each surface node, yielding an output $P \in \mathbb{R}^{N}$.

To model complex spatial interactions, AdaField leverages vector self-attention to extract pointwise geometric features, with each attention block followed by a Flow-Conditioned Adapter (FCA) to integrate flow-specific information. We adopt a U-Net-style \cite{ronneberger2015unetconvolutionalnetworksbiomedical} architecture to support dense pointwise prediction. At each stage, after feature extraction with vector self-attention and FCA, AdaField introduce a semantic aggregation module for downsampling. Symmetrically, the decoder progressively upsamples point features via k-nearest neighbor (kNN) interpolation and incorporates skip-connected features from the corresponding encoder layers. The final pressure prediction is produced by a linear output layer applied to the recovered high-resolution point cloud.

\subsection{Preliminary: Vector Self-Attention for Point Cloud}

AdaField adopts vector self-attention, as introduced in Point Transformer \cite{zhao2021pointtransformer}, to enhance its capability in capturing local geometric structures from point cloud data. Each point transformer layer takes as input a set of features and their corresponding coordinates, denoted as $\mathrm{input} = (x, p)$, where $x \in \mathbb{R}^{N \times D}$ represents the point features and $p \in \mathbb{R}^{N \times 3}$ represents the 3D coordinates.

The attention output $y_i$ for the $i^{\text{th}}$ point is computed as:

\begin{equation}
	\begin{aligned}
		& \delta_{ij} = \mathcal{P}(p_i - p_j),\\
		& q_i, k_i, v_i = \mathcal{Q}(x_i), \mathcal{K}(x_i), \mathcal{V}(x_i),\\
		& y_i = \sum_{x_j \in N(x_i)} \mathrm{Softmax}(\gamma(q_i - k_j + \delta_{ij})) \odot (v_j + \delta_{ij}),
	\end{aligned}
\end{equation}
where $\mathcal{Q}, \mathcal{K}, \mathcal{V}$ are linear projection functions, $\mathcal{P}$ and $\gamma$ are two-layer MLPs, and $N_k(x_i)$ denotes the set of k-nearest neighbors of point $x_i$.
To further enrich the representation, we follow the standard Transformer design by applying a residual connection and a feed-forward network (FFN):

\begin{equation}
	\begin{aligned}
		& x_i \leftarrow x_i + y_i,\
		& x_i \leftarrow x_i + \mathrm{FFN}(x_i).
	\end{aligned}
\end{equation}

\subsection{Semantic Aggregation Module}

Typical point cloud downsampling methods use farthest point sampling (FPS) \cite{623193} to select a subset of points, followed by feature aggregation within the k-nearest neighbors (kNN) of the selected points. However, such approaches rely solely on spatial proximity and do not fully leverage the underlying geometric similarity between local structures, which we consider crucial for generalization across diverse vehicle shapes.

To address this, we introduce a semantic aggregation module, which is adapted from Slot Attention \cite{locatello2020object}, originally developed for object discovery \cite{villavasquez2024unsupervisedobjectdiscoverycomprehensive}. Unlike conventional distance-based grouping, the semantic aggregation module groups points based on learnable feature similarity, allowing the network to capture semantic level structures from the input geometry.

Formally, given input $\mathrm{input} = (x, p)$, where $x \in \mathbb{R}^{N \times D}$ denotes the point features and $p \in \mathbb{R}^{N \times 3}$ the coordinates, we use FPS to select $K$ initial aggregation center $S = (x_s, p_s)$, where $x_s \in \mathbb{R}^{K \times C}$ and $p_s \in \mathbb{R}^{K \times 3}$. Then $x_s$ are updated over several iterations using the following formulation to integrate information from other points:

\begin{equation}
	\begin{aligned}
		& \delta_{ij} = \mathcal{P}_s(p_{s,i} - p_j),\\
		& q_i, k_i, v_i = \mathcal{Q}_s(x_{s,i}), \mathcal{K}_s(x_i), \mathcal{V}_s(x_i),\\
		& y_{s,i} = \sum_{x_j \in N_k(x_{s,i})} \mathrm{Softmax}(\gamma_s(q_i - k_j + \delta_{ij})) \odot (v_j + \delta_{ij}),\\
		&x_s \leftarrow \mathrm{GRU}(x_s, y_s),\\
		&x_s \leftarrow x_s + \mathrm{FFN}(x_s),\\
	\end{aligned}
\end{equation}

Here, $\mathcal{Q}_s, \mathcal{K}_s, \mathcal{V}s$ are linear projection functions, $\mathcal{P}s$ and $\gamma_s$ are two-layer MLPs, and $N_k(x_{s,i})$ denotes the k-nearest neighbors of slot $x_{s,i}$ within the point set $x$.

\subsection{Flow-Conditioned Adapter}

In the analysis of surface pressure, different types of transportation systems usually require consideration of different input flow conditions. For instance, the surface pressure field of cars and trains is typically influenced by the driving speed and crosswind speed \cite{qiu2025compressed}. In contrast, for aircraft, the flow conditions are usually described by the Mach number (Ma) and angle of attack (AoA) \cite{zhang2024prediction}. This hinders the model's universality in different aerodynamic scenarios. 

To address this issue, we propose the Flow-Conditioned Adapter (FCA), which independently learns representations for different flow conditions. FCA is added after each vector self-attention layer. In each FCA, the input point cloud features $x$ is first projected to a lower dimension. Then, a set of scale $\sigma$ and bias $\mu$ is generated with the flow condition $C$ through a two-layer MLP to modulate the projected features. Finally, the features are projected back to the initial dimension and added to the original feature through a residual connection. The projections are accomplished through a linear layer, a LayerNorm, and a GELU activation function. Formally,

\begin{equation}
	\begin{aligned}
		& \sigma, \mu = \mathrm{MLP}(C), \\
		& y = \mathcal{P}_{\mathrm{out}}((\mathcal{P}_{\mathrm{in}}(x) + \mu) \odot \sigma),\\
		& x \leftarrow x + y,
	\end{aligned}
\end{equation}

where $\mathcal{P}_{\mathrm{in}}$ and $\mathcal{P}_{\mathrm{out}}$ are projections comsist of linear layer, LayerNorm, and GELU activation.

\subsection{Physics-Informed Data Augmentation}

A major obstacle to achieving generalizable surface pressure field prediction models across different engineering domains is the substantial variation in object scale and velocity. For example, a typical car measures approximately 5 meters in length and usually travels at speeds below 40 m/s, whereas high-speed trains can extend tens of meters and reach speeds close to 100 m/s. Consequently, data from high-speed trains represent a completely unseen distribution for models trained solely on car datasets, posing significant challenges to generalization.

To overcome this, we propose a Physics-Informed Data Augmentation (PIDA) technique grounded in the steady-state incompressible Navier-Stokes equations \cite{constantin1988navier}. This approach systematically scales existing data in both scale and velocity, thereby broadening the effective training distribution and improving model adaptability. Formally, we state the following theorem:

\begin{theorem}
	Let $G \in \mathbb{R}^{N \times 3}$ represent the geometric shape of a vehicle, $v \in \mathbb{R}$ the traveling speed, and $C_p \in \mathbb{R}^{N}$ the dimensionless surface pressure coefficient. Under steady-state incompressible flow conditions, the geometric shape $G' = G/c$ traveling at speed $v' = cv$ will have the same surface pressure coefficient $C_p$.
\end{theorem}

The proof is provided in the supplementary material. By varying the scaling factor $c$, we can generate surface pressure fields corresponding to larger objects moving more slowly or smaller objects moving faster, effectively expanding the data distribution. In practice, we define a range $[c_{\min}, c_{\max}]$ and uniformly sample $c$ within this range to augment the size and speed of training samples.

\begin{figure*}[t]
	\centering
	\includegraphics[width=2.1\columnwidth]{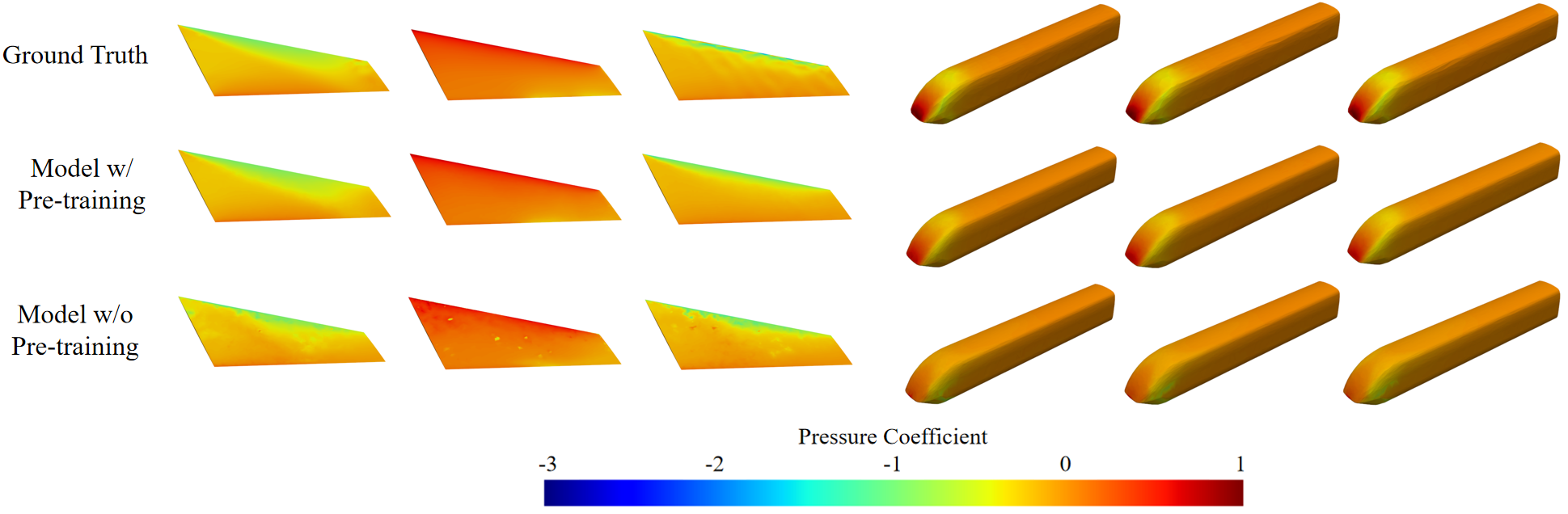}
	\caption{Pre-trained models for geometry generalization. When applying the model to maglev trains and aircraft wings, the model with and without DrivAerNet++ pre-training makes a huge difference. In the aircraft wing dataset, the results provided by the pre-trained model are more natural, with smooth and stable pressure changes. As for maglev trains, model without pre-training gives almost a completely uniform distribution over the entire surface. This indicates that when data in a certain small domain is scarce, it is insufficient to train a model from scratch that can generalize well to unknown geometries, while pre-trained model from DrivAerNet++ benefits the generalization ability.}
	\label{fig: comparison_generalization}
\end{figure*}

\section{Experiments}

In widely studied domains such as the automotive aerodynamics, large-scale datasets are publicly available. However, in specialized fields like trains or aircraft, only limited data can be obtained through computational fluid dynamics (CFD) simulations. Such scarce data make it challenging to train a generalizable model from scratch. Therefore, transferring pre-trained models from related fields to these specialized domains is critical.

In this section, we first pre-train AdaField on the automotive dataset DrivAerNet++ for the surface pressure field prediction task and benchmark its performance against current SOTA models to demonstrate its performance superiority. Subsequently, we introduce datasets for trains and aircraft wings, and fine-tune AdaField on these target domains using only a limited number of samples. We compare the results with models trained without pre-training to validate that AdaField effectively captures general aerodynamic knowledge during pre-training and generalizes well to novel domains.

\begin{table}[t]
	\small
	\begin{center}

		\setlength{\tabcolsep}{1mm}
		\begin{tabular}{l c c c c c} 
			\toprule[2pt]
			\makebox[0.14\textwidth][c]{\multirow{3}{*}{\textbf{Model}}}  & 
			\multicolumn{5}{c}{\makebox[0.05\textwidth][c]{DrivAerNet++}} \\
			\cmidrule(r){2-6}
			& $\downarrow$MSE & $\downarrow$MAE & $\downarrow$MaxAE & $\downarrow$RelL2 & $\downarrow$RelL1 \\
			& ($\times 10^{-2}$) & ($\times 10^{-1}$) &  & (\%) & (\%) \\
			\hline
			RegDGCNN \shortcite{elrefaie2024drivaernet} 	& 8.29 & 1.61 & 10.81 & 27.72 & 26.21 \\
			Transolver \shortcite{Transolver} 	& 7.15 & 1.41 & 7.12  & 23.87 & 22.57 \\
			FigConvNet \shortcite{FIGConvNet} 	& \underline{4.99} & \underline{1.22} & 6.55  & 20.86 & 21.12 \\
			TripNet \shortcite{TripNet} 	& 5.14 & 1.25 & \underline{6.35}  & \underline{20.05} & \underline{20.93} \\
			\hline
			\textbf{AdaField} (ours) & \textbf{4.58} & \textbf{1.05} & \textbf{6.22} & \textbf{19.81} & \textbf{17.28} \\
			\bottomrule[2pt]
		\end{tabular}
		\caption{Model performance comparison on DrivAerNet++.}
		
		\label{tab: drivaernet++comparison}
	\end{center}
	
\end{table}

\begin{figure*}[t]
	\centering
	\includegraphics[width=2.1\columnwidth]{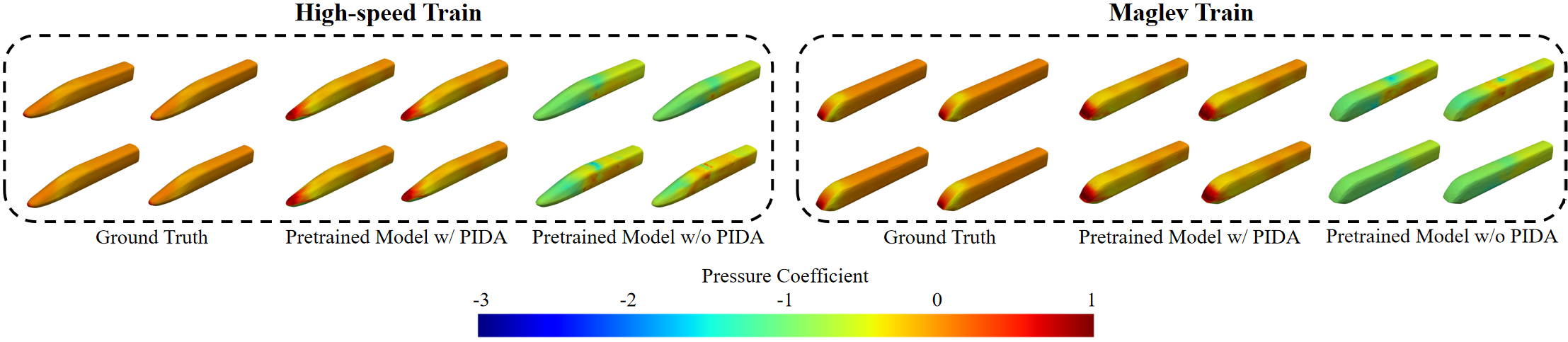}
	\caption{Applying the pre-trained model directly to the train scenario \textbf{without fine-tuning}. The model with PIDA predicts pressure distributions that closely match the ground truth. In contrast, the model without PIDA produces negative pressure across the entire surface, even including at the front, highlighting its lack of physical consistency.}
	\label{fig: comparison_PIDA}
\end{figure*}

\subsection{Datasets}

\textbf{DrivAerNet++} \cite{DrivAerNet++}: DrivAerNet++ is a publicly available large-scale, high-fidelity dataset tailored for learning-based aerodynamic analysis of vehicles. All data are generated from high-resolution CFD simulations using steady-state incompressible flow solvers. Each sample captures the 3D surface pressure field of a vehicle traveling at 30 m/s. In total, DrivAerNet++ contains over 8,000 samples of car model surface point clouds with corresponding pressure fields, each consisting of more than 500,000 surface points. This rich dataset supports various aerodynamic prediction tasks, including lift and drag estimation, as well as dense parameter predictions such as surface pressure and wall shear stress. Our focus in this work is primarily on the surface pressure prediction task.

\noindent \textbf{Train dataset} \cite{qiu2025compressed}: The train dataset comprises geometries of one high-speed train and two maglev trains, with CFD simulations conducted under multiple flow conditions. Specifically, flow conditions are characterized by the train’s traveling speed and the lateral wind speed perpendicular to the travel direction. For the high-speed train, speeds range from 270 km/h to 350 km/h in 10 km/h increments, and lateral wind speeds range from 0 m/s to 15 m/s in 5 m/s increments, yielding 36 samples. For each maglev train geometry, speeds vary from 70 m/s to 100 m/s in 10 m/s increments, and lateral wind speeds range from 0 m/s to 20 m/s in 5 m/s increments, producing 20 samples per geometry.

\noindent \textbf{Aircraft wing dataset} \cite{Wang_Zhang_Liu_Zhao_Lin_Chen_2025}: This dataset includes 10 aircraft wing geometries. Flow conditions are described by Mach number and angle of attack. For each geometry, Mach numbers span from 0.2 to 0.8 in 0.2 increments, and angles of attack vary from –15° to 15° in 5° increments, resulting in 28 samples per geometry.

\subsection{Metrics}

We adopt the evaluation metrics used in the DrivAerNet++ leaderboard for a comprehensive assessment of model performance. These metrics include Mean Squared Error (MSE), Mean Absolute Error (MAE), Max Absolute Error (Max AE), Relative L2 error (RelL2), and Relative L1 error (RelL1). \textbf{Lower is better} for all the metrics.

\begin{table}[t]
	\small
	\begin{center}

		\begin{tabular}{c c c c} 
			\toprule[2pt]
			\multirow{2}{*}{\textbf{Dataset}} &
			\multirow{2}{*}{\textbf{Pre-trained}} & $\downarrow$MSE & $\downarrow$MAE \\
			&& ($\times 10^{-2}$) & ($\times 10^{-1}$) \\
			\hline
			\multirow{2}{*}{\textbf{Aircraft wing}} 
			&            & 16.63 & 2.18 \\
			& \checkmark & \textbf{11.46} & \textbf{1.52} \\
			\hline
			\hline
			\multirow{2}{*}{\textbf{Train}}	  
			&            & 1.79 & 0.51 \\
			& \checkmark & \textbf{0.99} & \textbf{0.38} \\
			
			\bottomrule[2pt]
		\end{tabular}
		\caption{Comparison between fine-tuned and trained-from-scratch model. Although Train and Aircraft Wing scenarios are largely different from cars, the error is still significantly reduced when pre-trained on DrivAerNet++.}
		
		\label{tab: few_shot_adaptation}
	\end{center}
	
\end{table}

\subsection{DrivAerNet++ Benchmark Comparison}

\textbf{Setup:} Following prior works \cite{FIGConvNet,TripNet}, we select a random subset of 32,768 points sampled from the full point cloud for both training and testing. The train-test split adheres to the official partition provided in the DrivAerNet++ GitHub repository. All pressure values are normalized by subtracting the mean (-94.5) and dividing the standard deviation (117.25).

Flow conditions comprise the driving speed $v$ and the crosswind speed $w$. Since DrivAerNet++ does not simulate crosswind, $w$ is fixed at zero but retained for consistency with the subsequent train dataset. We expand the driving speed and car scale ranges using PIDA. Specifically, $c_{min}$ and $c_{max}$ are set to 0.2 and 3.2, extending the speed range to 6–96 m/s and vehicle size to 1.5–25 m. During testing, no augmentation is applied, and original data are used.

\noindent \textbf{Results:} Results are summarized in Table \ref{tab: drivaernet++comparison}. AdaField consistently outperforms all baselines. In particular, it achieves the lowest Mean Squared Error (MSE) of 4.58 $\times 10^{-2}$, improving upon the previous best FigConvNet (4.99) by approximately 10\%. Regarding Mean Absolute Error (MAE), AdaField attains 1.05 $\times 10^{-1}$, more than 15\% lower than the second-best competitor. Furthermore, it achieves the best Max Absolute Error (Max AE), indicating improved robustness against outliers. These results establish AdaField as the new SOTA for surface pressure prediction.

\subsection{Generalizing to Novel Domains}

\textbf{Setup:} We fine-tune the model pre-trained on DrivAerNet++ with limited data to transfer it to novel domains, including the train and aircraft wing datasets. During fine-tuning, only the parameters of the Flow-Conditioned Adapter (FCA) are updated. For the train dataset, training is performed on high-speed train and one of the maglev train geometry, with testing conducted on another maglev train. For the aircraft wing dataset, the model is trained on data from five wing geometries and tested on the remaining five.
To verify the benefit of pre-training, we compare this fine-tuned model against one trained from scratch, where all parameters are trained on the target data.

\noindent \textbf{Results:} As shown in Table \ref{tab: few_shot_adaptation}, the fine-tuned model significantly outperforms the model trained from scratch, reducing prediction errors by about one-third on both datasets. For the train dataset, visualization in Figure \ref{fig: comparison_generalization} highlight this contrast: for maglev train, the surface pressure typically shows positive value at the front, negative value at the junction between the front and the body, and near-zero values along the remaining body. The fine-tuned model gives similar distribution, whereas the trained-from-scratch model predicts a pressure distribution that fails to reflect the sharp pressure variations at the front of the train. For the aircraft wing dataset, the fine-tuned model also substantially lowers prediction errors. In Figure \ref{fig: comparison_generalization}, the prediction results of the fine-tuned model are more consistent with the ground truth, while its counterpart's prediction results are unstable and there emerges abnormal patches.
These results demonstrate that despite notable differences between pre-training and target data, AdaField successfully captures transferable aerodynamic knowledge and enables effective generalization with minimal target-domain data.

\begin{table}[t]
	\small
	\begin{center}

		\begin{tabular}{c c c c c} 
			\toprule[2pt]
			\multirow{2}{*}{\textbf{Dataset}} &
			\multirow{2}{*}{\textbf{SAPT}}  &
			\multirow{2}{*}{\textbf{PIDA}} &
			$\downarrow$MSE & $\downarrow$MAE \\
			&&& ($\times 10^{-2}$) & ($\times 10^{-1}$) \\
			\hline
			\multirow{2}{*}{\textbf{DrivAerNet++}} 
			& & \checkmark & 5.22 & 1.14 \\
			& \checkmark & \checkmark & \textbf{4.58} & \textbf{1.05} \\
			\hline
			\hline
			\multirow{3}{*}{\textbf{Aircraft Wing}}	 
			& & \checkmark & 13.05 & 1.79 \\
			& \checkmark & & 12.95 & 1.91 \\
			\cline{2-5}
			& \checkmark & \checkmark & \textbf{11.46} & \textbf{1.52} \\
			\hline
			\hline
			\multirow{3}{*}{\textbf{Train}}	 
			&& \checkmark & 1.15 & 0.44 \\
			& \checkmark & & 1.35 & 0.61 \\
			\cline{2-5}
			& \checkmark & \checkmark & \textbf{0.99} & \textbf{0.38} \\
			\bottomrule[2pt]
		\end{tabular}
		\caption{Ablation Studies. We consider SAPT, PIDA and ablate one component in each row to show the contribution of each component to final performance.}
		
		\label{tab: ablation}
	\end{center}
	
\end{table}

\subsection{Ablation Studies and Analysis}

This work tackles the challenges of limited data in surface pressure field prediction through multiple improvements at both the model and data, including SAPT that integrating a semantic aggregation module and vector self-attention to enhance feature representation, FCA that enables adaptation to unseen flow conditions and parameter-efficient fine-tuning, as well as PIDA that broadens the training data to boost generalization to unseen data. Here we perform ablation studies and analyses to validate the contributions of each component.

\subsubsection{Semantic Aggregation Module}

The semantic aggregation module acts as a replacement for the FPS + KNN downsampling strategy which is commonly used in point cloud processing. The comparison is shown in Table \ref{tab: ablation}, where the models without a checkmark at SAPT indicate using conventional FPS + KNN downsampling. For all the datasets, incorporating semantic aggregation module brings about remarkable performance boost, yielding approximately 10\% to 20\% error reduction for all the datasets.

\subsubsection{Physics-Informed Data Augmentation}

PIDA enables AdaField to generalize across a broader range of velocity and scales, thereby improving its robustness and adaptability in unseen scenarios. As shown in Figure~\ref{fig: comparison_PIDA}, we visualize the surface pressure field predicted by models with and without PIDA when directly applied to the train datasets \textbf{without fine-tuning}.
The model incorporating PIDA produces surface pressure distributions that closely resemble the ground truth, capturing key physical patterns such as positive pressure at the front, negative pressure at the transition region, and near-zero pressure along the main body. In contrast, the model without PIDA fails to predict pressure distributions that are inconsistent with physical common sense, e.g., giving a negative pressure at the front of the train.

These results visually highlight the effectiveness of PIDA in learning physically consistent and transferable knowledge, enabling the model to generalize to novel transportation systems. This also forms the basis for the better performance of models with PIDA after fine-tuning, as shown in Table~\ref{tab: ablation}, where the error of the model with PIDA after fine-tuning reduces by about 15\% for Aircraft Wing and one-third for Train, compared with that without PIDA.

\subsubsection{Flow-Conditioned Adapter}

In addition to adapting the model to different flow conditions, FCA also enables the model to perform parameter-efficient fine-tuning, thereby achieving few-shot transfer learning. To verify the model's efficiency in utilizing samples, we further reduce the amount of data used for training, and also verify the most extreme case where fine-tuning using only one sample. 

\begin{figure}[t]
	\centering
	\includegraphics[width=1\columnwidth]{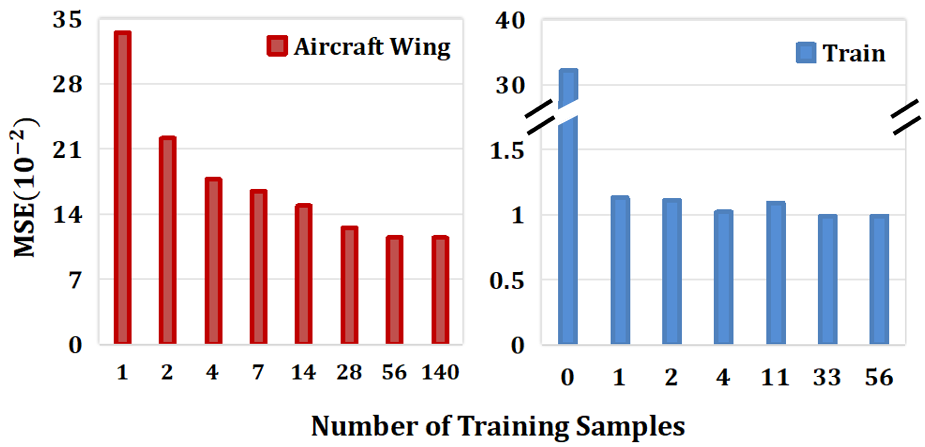}
	\caption{The performance trend of the model when reducing the number of samples for fine-tuning.}
	\label{fig: num_sample}
\end{figure}

Figure \ref{fig: num_sample} gives the detailed performance trend when using different number of samples. For the train dataset, despite the reduced amount of data, the models do not show significant performance degradation. Even if only one sample is used, the error can be reduced to a relatively low level, which may be related to the fact that the pressure distribution among different train data remains similar. For aircraft wing data, due to the influence of the angle of attack, the pressure distribution varies significantly across different samples, thus requiring more data for fine-tuning. Nevertheless, when using about 28 samples, 20\% of all the samples, the prediction error of the model can still get close to the model trained with entire 140 samples.

\section{Conclusion}

The surface pressure field plays a vital role in shaping designs that reduce energy consumption and enhance operational stability. Neural networks have emerged as an effective alternative to traditional computationally intensive methods like CFD simulations for surface pressure prediction. However, many subdomains of fluid dynamics suffer from a scarcity of high-fidelity data, limiting the widespread adoption of data-driven approaches.

We focus on the solution to pre-train models on publicly available large-scale datasets and transfer them to data-scarce specialized domains. In this paper, we validate this approach by designing AdaField, a framework that incorporates Semantic Aggregation Point Transformer (SAPT), a high-performance network for processing point clouds, Flow-Conditioned Adapter (FCA) that enables adaptation to varying flow conditions through parameter-efficient fine-tuning, and Physics-Informed Data Augmentation (PIDA) to expand the data distribution and enhance generalization.

Our experiments demonstrate that the proposed components not only achieve SOTA performance on the public DrivAerNet++ dataset, but also enable effective transferring to aircraft and train domains with markedly reduced errors compared to training from scratch. Our results highlight the potential of neural networks to unify flow field representations across diverse domains and serve as universal surface pressure prediction models, providing support for aerodynamic optimization in various transportation systems.

\section{Acknowledgments}

This work was supported by the National Key R\&D Program of China (No. 2025ZD0122000), the Strategic Priority Research Program of Chinese Academy of Sciences under Grant XDA0480103, Chinese National Natural Science Foundation 92570119, and the China National Railway Group Science and Technology Program (Grant No. N2024J040).

This paper is finished under the supervision of Prof. Zhen Lei.

\bibliography{aaai2026}

\end{document}